# Toward an anthropocentric approach for hybrid control architectures: case of a furniture factory

Etienne Valette [1,2,3](✉), Hind Bril El-Haouzi[1,2], Guillaume Demesure[1,2], Vincent Boucinha[3]

[1]Université de Lorraine, CRAN, UMR 7039, Campus Sciences, BP 70239, 54506 Vandœuvre-lès-Nancy cedex, France
[2]CNRS, CRAN, UMR7039, France
(e-mail: etienne.valette5@etu.univ-lorraine.fr; {hind.el-haouzi; guillaume.demesure}@univ-lorraine.fr)
[3]Parisot Meubles, 15 avenue Jacques Parisot, 70800 Saint-Loup-sur-Semouse
(e-mail: {evalette; vboucinha}@parisot.com)

**Abstract.** Typology of goods and services' consumption has changed. In order to adapt to this change, it is relevant for a company to turn toward new ways of production and management. Slowly, the concept of industry 4.0 starts to set up in manufacturing companies. Research on hybrid control systems favours achieving automated and flexible production system through "*jidoka*" (or *autonomation*) and *Just In Time* principles. Still, studies stay mainly techno-centred rather than anthropocentric. *Parisot* company suffers today of a lack of reactivity to its market and of a production tool maladjusted to customer's consumption habits. This article aims to sum up the company's current situation, and to introduce the thesis project that intends to back it through its economic, technological and sociological transition toward a flexible, adaptive and sustainable human-centered manufacturing system.

## 1 Introduction

Pine II [1] was one of the very first to identify and enunciate the concept of mass customization. Ten years ago, Klein stated about this concept "Typology of consumption of goods and services has evolved over the last few years, especially with the emergence of e-commerce. [...] the trend is now to personalize them, implying an increase of diversity and of end customer requirements in terms of costs, quality, features and delivery time." [2]. That assertion has been proved to be true, especially in the furniture industry, which Parisot, Saint-Loup's Production Unit (PSL) is part of. As a matter of facts, Klein wrote this as part of his thesis within this company. Customers need complex, customizable and quality products, available by e-commerce at the lowest price and with the shortest delivery time. This is no more a trend: this is the new way of consuming manufactured goods. Manufacturers' production tools, like PSL's one, were and are still designed to produce large batches of undifferentiated products. New ways of consuming imply new ways of producing for manufacturers. Still, like PSL, many manufacturers haven't changed their production methods yet.

Zimmermann have identified the necessity to master the material and information flows and the importance of keeping these flows flexible and adaptable to the market demand as being the main challenges of this new consuming habits [3]. In fact, these challenges can be considered as secondary for many companies: the main challenge would be to adapt an old complex and rigid pre-existing industrial tool to new production methods. Moreover, what is rarely considered in these situations is the strong tendency the different actors have to resist to any changes, even if these changes are of vital necessity.

A company is represented as a moral person and considered as a complex system, composed by humans, tools and data. Koestler defines a holon as something both whole and part, and where parts keep properties of global system, despite their independence [4]. If a company is both a whole entity (moral person) and the sum of its parts (humans, tools and data), then it is to be considered as a holonic system, simultaneously parts and whole. This paradigm can easily be linked to the complex apprehension of the universe praised by Morin [5]. Then, the socio-ecological transition that should be undertaken by mankind would be associated to the necessary anthropocentric technical and economic transition of the industry [6]. What Morin assess as unavoidable to achieve socio-ecological transition, is the need of "reliance1" between each part of the universe.

This philosophical concept is surprisingly close to the need of breaking the knowledge and responsibility partitioning between the services and actors of the material and data flows in a company. This partitioning finds its origin in the strong specialization of the flows' actors, which are grouped by in separated offices. Since services are physically separated, effective and efficient communication becomes uneasy. Responsibility is rejected on upstream and downstream services, problems are solved locally without consulting other actors and communication slowly turns to permanent conflict. Data and information becomes valuable pressure levers on other services, and get jealously protected. Hence, data reliability decrease and the whole flow weaken. Instead of developing a team spirit, knowledge's partitioning generates a competitive spirit. Huybens explains these phenomena trough the psycho-sociology of restricted groups approach [8].

---

[1] The word is based on the French word "relier", meaning *connect*. This concept was first studied from 1975 to 1985 by Bolle De Bal [7]. (Morin, 2014) use this term to describe a need to connect to each other's and to the universe, in order to achieve the socio-ecological transition.

This approach and issues are really interesting to study since they are rarely taken into consideration. If Klein, Zimmermann and many others consider the human factor in their works, this element stays minimal in regard to the importance given to the technical studies. What is studied is the impact of the human **on** the process. But what could be interesting to study would be the interactions **between** the human **and** the process and even the interactions **between** the human **and** the process **as parts of** a larger entity, such as the company.

Thus, we can assess that PSL is suffering from a great lag in its adaptation to its market. The adaptation of the company to the new consumption ways will occur through a complex transition, as much human as technological and economic. This article brings a complex vision of the current situation and suggests research leads that should be depth as part of a CIFRE2 thesis. The present paper will now quickly introduce **the economic transition** that is today operated by the company, and then will focus on the issues and challenges that will be raised by **the technological and organizational transition**. The fourth part of this paper will focus a bit more on the **sociological transition** and on human challenges. Finally, the conclusion will quickly sum the purpose of this article up and introduce the outlooks that could be considered.

## 2 Initiating economic transition

The company has its own New Products Development (NPD) process. This process goes through services like trading, design, marketing, innovation, technical and Industrialization studies and prototypes workshop. This process has two entries. In one hand, a client demand can trigger the development of a new product (or product range). In the other hand, the company occasionally suggests to its clients a product range complement or can spontaneously decide to renew a part of its own collection (Fig. 1).

---

[2] CIFRE : Industrial Convention of Formation through REsearch

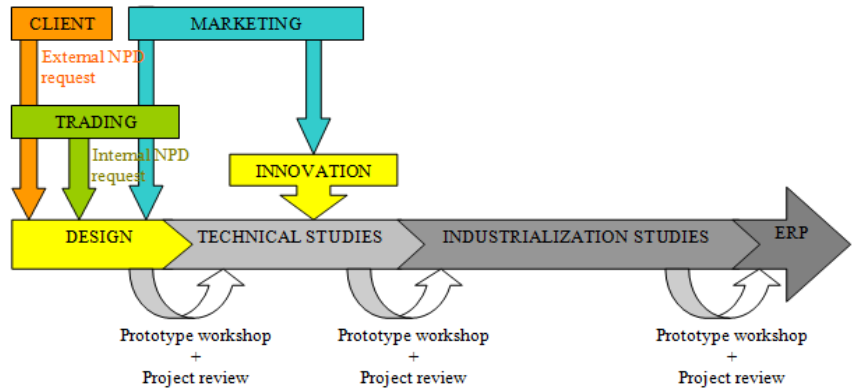

**Figure 1 :** PSL's New Products Development process

Today, around 80% of sales are still signed with specialized (60%) and DIY (20%) superstores, and concern large production batches (more than 1.000 products). Around 25% of the sales are made for export (European and worldwide).

Today, PSL is strongly attached to its great historical clients (specialized superstores like *Conforama*, *But* and *Fly*). This dependence is a real threat for the company since its sells are tributaries of superstores' ones. PSL lacks effective negotiation levers towards its clients and have no satisfying fallback solution in case of loss of one of them. The company almost faced this situation recently with the economic difficulties that has touched *Conforama*'s headquarter in December 2017 [9].

Since specialized superstores' demand became increasingly unpredictable and uncertain, the company tries to reinforce its sells with DIY superstores. Unfortunately, this can only be a provisional solution. According to the Prospective and Furnishing Studies' Institute (IPEA) "*DIY brands are increasing their development on furniture's trade, but global sales are struggling despite the recovery in real estate dynamics. The furniture's trade seems to be a source of growth for them, but for now, the creation of a specific identity on the market is cruelly lacking to each one of them*" [10].

Mail-order selling and home delivery are not new concepts in the company, as part of the activity for more than twenty years. These are today reaching around 20% of the sales (in terms of volume and turnover). Regarding e-commerce, IPEA, FNAEM3, and UA4 's studies assess "*Regarding the other circuits of the market, growth is also on time on e-commerce circuit both among historical mail-ordering players and pure-players. Many players are speeding up on the furniture's trade as shown by the strong double-digit progressions displayed by some.*" [11]. Therefore, PSL turned to the reinforcement of its internet commerce and exports (Fig. 2): for 2018, the sales objectives are 35% for export and 25% for e-commerce.

---

[3] FNAEM : French furniture and home equipement's trading federation
[4] UA : National Union of french furniture's industries

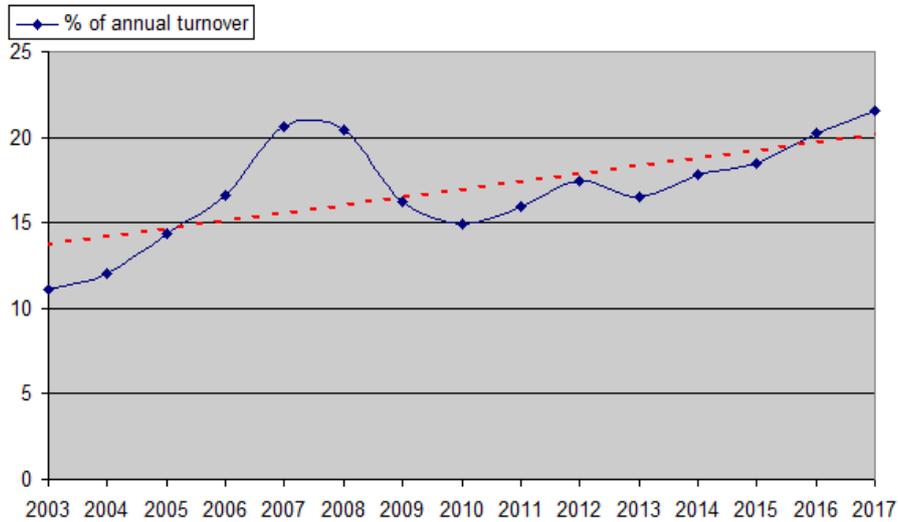

**Figure 2:** Evolution of PSL's annual exports (*Diane* export)

The economic transition is already in progress. Still, the gait is slow, hesitating and iterative. Many mistakes are today accepted, the first one being the lack of a global vision and approach. Nowadays, the company needs to speed up this transition, especially through the upgrade of its production methods and tool.

## 3  Technological and organizational transition: launching production tools' upgrade

PSL's production tool is the inheritance of eighty years of industrial development. The industrial site was originally organized in Autonomous Production Units (APU). The most recent unit, UGV, was inaugurated in 1996. Since, these units became less and less autonomous, as services like technical and industrialization studies were merged and centralized. The only unit to be considered as APU remaining is PMK (Fig. 3).

On this map, the red circle points the offices, the three green circles points the main production units and the two blue circles are for support production units. The MAF is the centralized storehouse for all products that are not particle boards or packaging. UDC is the main storage, picking and shipping unit. SAV stands for After Sales Service.

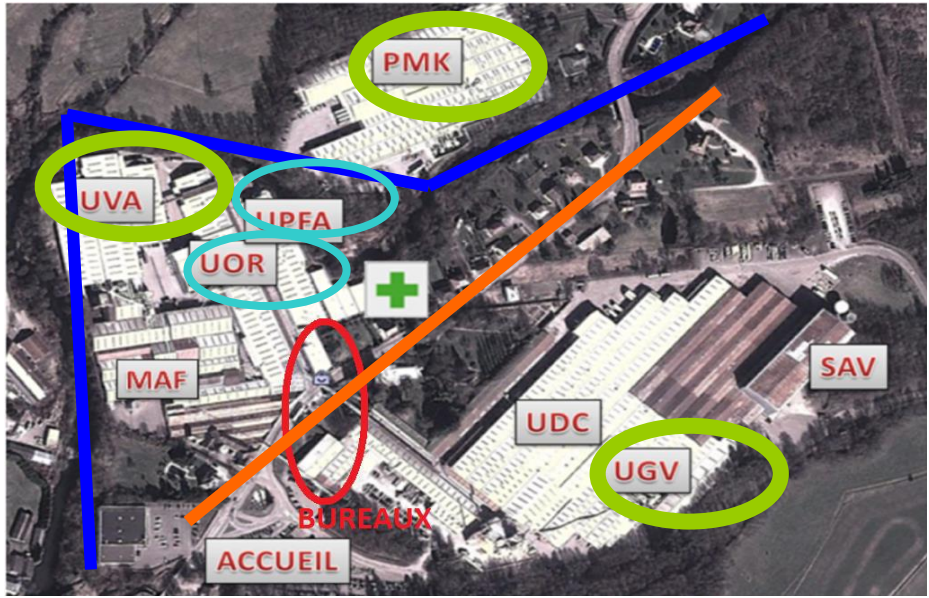

**Figure 3:** PSL's aerial view

This industrial site has two geographical particularities. The first one is the road (orange line) and the river (blue line) which split the site in three while separating the main production units. This leads to rather complex internal logistics issues. Many trucks and tractors are running through the site to deliver raw and transformed products, hardware or various materials. The second one lies in the location of the unit PMK. This production unit is the fastest one, and was designed for the greatest series' production, but is implanted in flood zone.

Figure 4 is an update of Klein's mapping. It's interesting to superimpose this representation to the aerial view, in order to get a better understanding of the complexity of physical flows in the company today (Fig. 5). Despite Klein's work and efforts to develop and implement a system of active Kanban, none of these flows is today automated or integrated in a global management, leading to an insufficient global service rate. Worst, economic difficulties occurred in 2012 and almost led the company to the bankrupt. These difficulties had a serious impact on both global and local management. For the company focused mainly on its survival, many good practices and procedures were lost and production management turned anarchic on many levels. Today, the company only starts to define and put in place good practices and rigorous procedures again.

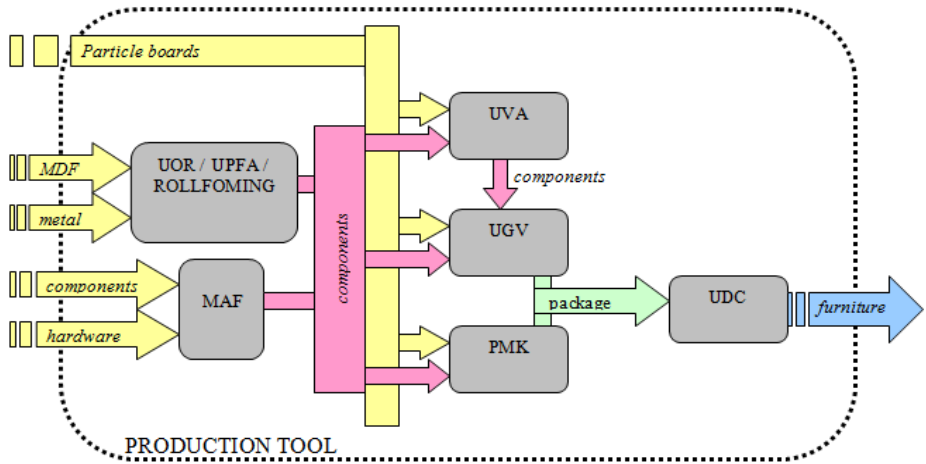

**Figure 4:** Main physical flows in PSL's production units

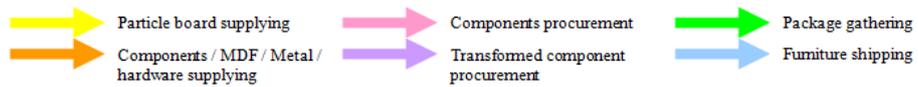

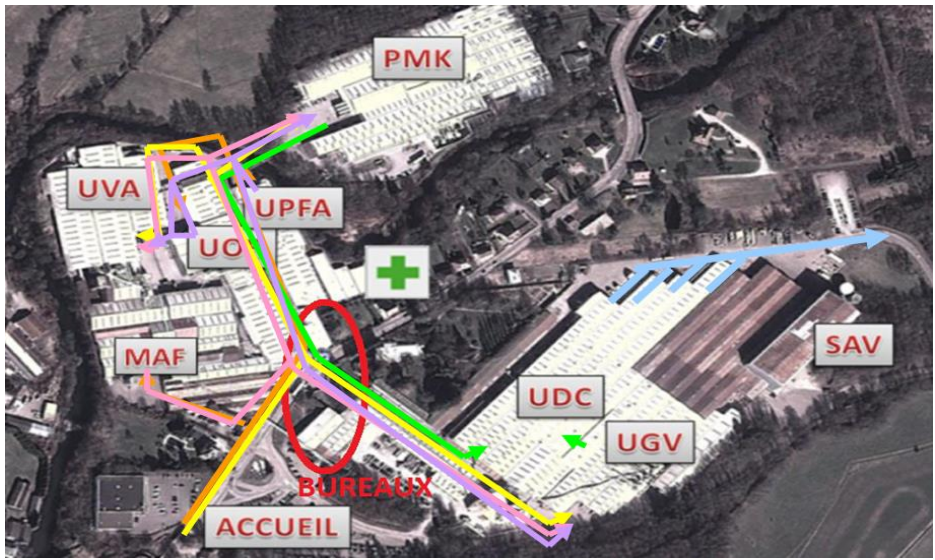

**Figure 5:** PSL's current physical flows

To limit and simplify these flows, a project of fusion between the units UVG and UVA has been evoked. This project would be a great opportunity to start the upgrade of the production tool. Still, many issues could be highlighted.

— **On the physical level**, mechatronics, data processing and information's transfer's technological breakthrough favours AGV[5] and drone's implanting and utilization [12]. Thanks to these breakthroughs the implanting could be quick and cheap and would bring a great flexibility in the production process. Some questions would then have to be considered. *What would be the best workshop configuration? Which of products, machines or both should be mobile? What roles and responsibilities should be given to mobile entities? How much would the system be flexible?*
— **On the informational** level, the main challenges would lie in the coordination between the mobile elements and between the local and global management systems. *Which level of intelligence should be given to mobile elements? What would be the nature of the decisions that would be delegated to them* [13]*?* The main objectives here are to guarantee the security of operators [14], the good functioning of the machines (mobile or not) and to prevent local machine taken decisions to be detrimental to the global industrial system [15].

These issues could mainly be solved by an intelligent use of the new ICTs[6] or artificial neural networks for machine learning and data mining (data mining could be used to refine the sales' previsions and then to improve the production schedule and stock level) [16].

## 4    Sociological transition: what about human aspects?

The most important element of a production system lies in its operators. The human being is the most flexible and "*agile*" element of the workshop. This assertion is particularly observable in PSL's productions units.

For example, to achieve the packaging step, the different pieces have to be spread on particular decks in order to be placed at the right place in the package. This implantation is designed by an industrialization studies' technician and conducted by a specialized operator. But if the designed implantation occurs to be improvable, the operator has the liberty to adapt it. This is the cause of many problems. If this liberty is allowed in order to maximize the local efficiency of the packaging line (in case of problem, the operator can solve it quickly, then inform the technician), its application is today full of drifts. Operators take the liberty to change the packaging plan without need, in order to increase the production rate, but to the detriment of the packages' quality. Since they only have an incomplete knowledge of the global process, what seems to be a good idea on their level turns out to be detrimental to the process. Once again emerges the problem of communication and of the risks presented by a local based management for the global management's performances. Lots of similar example could be indentified within the company.

---

[5] AGV: Automated guided vehicle
[6] ICT: Information and Communication Technology

Hence we can assess that PSL is already working with kind of hybrid system, in fact really near from **QRQC**[7] organization. This organization is based on a field approach (Gemba), with immediate reaction to specific issues, described with precise facts and data, by the people directly concerned. The manager comes as a support for the operator if it turns out that this one can not solve the problem by himself. The QRQC organization is a human-centered bottom-up approach of efficient problem solving.

Still, lack of rigor, formalization, and of continuous improvement of the system plus insufficient training, education and accountability of operators have led to a dysfunctional state where both global and local performances are degraded. An empowerment policy has been implemented in a system with dysfunctional information transformation and management system. The global entity has delegated some of its powers to its parts. Since the company is both part and whole, these powers are not lost, and should still operate. But what was forgotten was the principle of "reliance". Without it, parts could not accord to each others. PSL's example is the proof by the failure of the need of complex apprehension of a system on both global and local levels in order to successfully achieve its transition to Industry 4.0.

The concept of *Lean manufacturing* could be a way toward this achievement. It finds its origins in the *TPS*[8] system developed between the years 1950 and 1960 [17]. This concept is highly based on the humans for its two main pillars are *JIT*[9] and "*jidoka*" (or *autonomation*). The Jidoka is based on the "*andon*". The andons are visual cues by which the operator or the machine communicate and on which they can base their decisions. Short loops of identification, quality control and immediate reactions are then possible. They are the base of the global management. The JIT is based on a "*kanban*" system that is used to control local information loops, linked to steps of the physical production process. Thereby, the inertia of centralized information management systems can be avoided. Plus, global site management is prompted to work as much in the short term with ground issues as in the long term with fundamental causes analysis. Jidoka leads to JIT at a human level: according to Sakichi Toyoda[10], jidoka is to be considered as "*automation with a human touch*".

Jidoka leads to empowerment of the operators, by giving them decisional power and thus responsibilities. In the case of PSL, the empowerment could be even more important, since operators could contribute to the very thinking, creation, implantation and improvements of the system. But again, several questions already arise. Many employees are aged of 40 years old and more. These persons are mostly uneasy with new technologies 'usage. *How to deal with the question of age and formation in this transition? How will the question of the staff's competences evolutions be treated? What will be the new skills to search or develop to assure the sustainability, flexibility and continuous improvement of the company?*

---

[7] QRQC : Quick Response Quality Control
[8] TPS : Toyota Production System
[9] JIT : Just-In-Time
[10] Sakichi Toyoda (1867 – 1930) : founder of Toyota Industries, inventor of automatic power loom, implemented with Jidoka principles, creator of the 5Whys rule

In addition to that, a Kanban system will potentially suffers from difficulties to adjust to the variability of customers' requests. Plus, questions arise. *Which level of decisional power should be given to operators? How would operators efficiently interact with automated systems? How would the adequate level of information for taking decisions be given to operators?*

The use of ICT and artificial neural networks could to be viable solution parts to solve these issues. Anyway, what will be needed here are, in addition to technical expertise, strong managerial knowledge and skills doubled with both an accurate vision of the field reality and a global vision of tactical and even strategic objectives.

## 5      Conclusions and outlooks

*Yesterday*, a mass production tool with little visibility on the future was quite enough for a company to subsist. *Today*, ways of consumption have changed, and the industry struggles to answer to it. What is needed for *Tomorrow* is to think the transition in order to accord the industry to its present market and to prepare its future, especially through flexibility and adaptability (**Figure 6: economic, technological and sociological evolutions**).

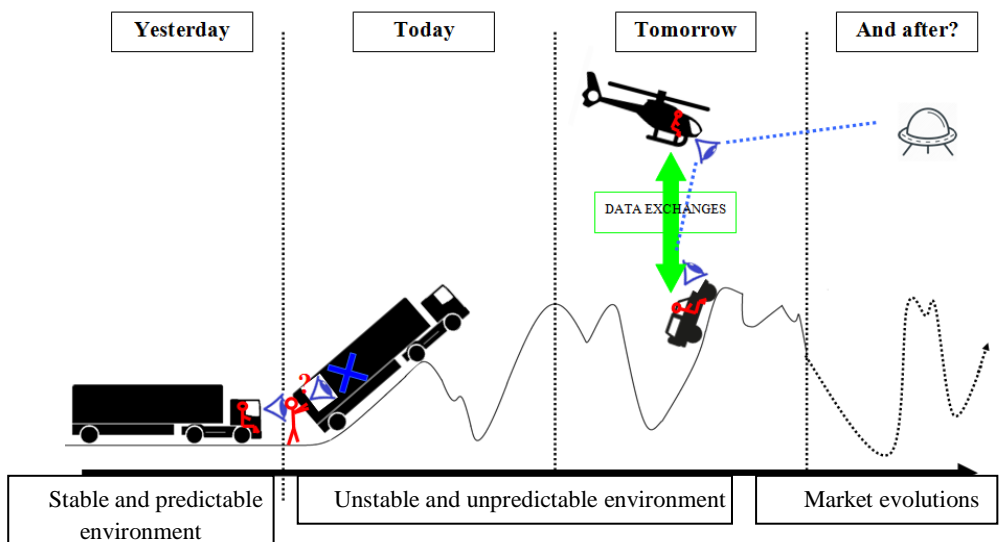

**Figure 6 : economic, technological and sociological evolutions**

The detailed analysis of the existing system, the conception and implanting and mastering of such a complex system is far from being an easy or common project. Some difficulties have been here identified, but many others will occur as the project will progress. Hybrids systems are no new concepts but globally stay at a state of experimental management. Robust decisions are very effective for recurrent issues, but lacks adaptability when confronted to unpredicted hazards.

The main objective of this paper was to bring a complex and representative vision of PSL's situation. In this article, some issues have been raised, and propositions have been made to partly solve them. PSL is obviously passing through an important transition phase. A better understanding of those issues and establishing of adequate solutions could be partly achieved through a CIFRE thesis within the company and in partnership with the CRAN[11].

---

[11] CRAN : Nancy's Research Centre in Automatic